\documentclass[%
 reprint,footinbib,
superscriptaddress,
 floatfix,
 amsmath,amssymb,
 aps,
prapp, 
longbibliography,
]{revtex4-2}

\usepackage{graphicx}  % needed for figures
\usepackage{dcolumn}   % needed for some tables
\usepackage{bm}        % for math
\usepackage{amssymb}   % for math
\usepackage{amsmath}
\usepackage{color}
\usepackage[colorlinks=true, pdfstartview=FitV, linkcolor=blue, citecolor=blue, urlcolor=blue]{hyperref} % enable links
\usepackage{epstopdf}
\usepackage{siunitx} % manages nice display of SI units and numbers
\usepackage{todonotes} % enables todo notes as a form of commenting
\usepackage{tabularx} 
\usepackage{xspace}
\usepackage{braket}
\usepackage{color}
\usepackage{setspace}
\usepackage{mdframed}
\usepackage{framed}
\usepackage{multirow}
\usepackage{setspace}
\usepackage[export]{adjustbox}
\usepackage{varwidth}
\usepackage{float}

\usepackage{lipsum}
\usepackage[version=3]{mhchem}
\usepackage{color, colortbl}
\definecolor{beaublue}{rgb}{0.74, 0.83, 0.9}

\newcommand{\mr}[1]{\mathrm{#1}}

\newcommand{\affilETH}{Laboratory for Solid State Physics, ETH Z\"{u}rich, 8093 Z\"urich, Switzerland.}
\newcommand{\affilCSEM}{Centre Suisse d'Electronique et de Microtechnique SA (CSEM), 2002 Neuchâtel, Switzerland.}

\begin{document}

%\preprint{}

%\preprint{}

\title{Soft-clamped silicon nitride string resonators at millikelvin temperatures}

\author{Thomas Gisler}
\affiliation{\affilETH}
\author{Mohamed Helal}
\affiliation{\affilETH}
\author{Deividas Sabonis}
\affiliation{\affilETH}
\author{Urs Grob}
\affiliation{\affilETH}
\author{Martin H\'eritier}
\affiliation{\affilETH}
\author{Christian L. Degen}
\affiliation{\affilETH}
\author{Amir H. Ghadimi}
\email[Corresponding author: ]{amir.ghadimi@csem.ch}
\affiliation{\affilCSEM}
\author{Alexander Eichler}
\email[Corresponding author: ]{eichlera@ethz.ch}
\affiliation{\affilETH}

%\date{\today} % It is always \today, today, but any date may be explicitly specified

\begin{abstract}
We demonstrate that soft-clamped silicon nitride strings with large aspect ratio can be operated at \si{\milli\kelvin} temperatures. The quality factors ($Q$) of two measured devices show consistent dependency on the cryostat temperature, with soft-clamped mechanical modes reaching $Q > 10^9$ at roughly \SI{46}{\milli\kelvin}. For low optical readout power, $Q$ is found to saturate, indicating good thermalization between the sample and the stage it is mounted on. Our best device exhibits a calculated force sensitivity of $\SI{9.6}{\zepto\newton\per\sqrt\hertz}$ and a thermal decoherence time of \SI{0.38}{\second}, which bode well for future applications such as nanomechanical force sensing.
\end{abstract}

\maketitle

Over the last years, string and membrane micromechanical resonators made from high-stress silicon nitride (\ce{Si3N4}) have established themselves as a powerful system for quantum engineering ~\cite{Verbridge_2008,Anetsberger_2010,Reetz_2019,Reinhardt_2016,Tsaturyan_2017,Ghadimi_2018,Beccari_2021_hierarchical}. Applications range from electro-opto-mechanical transduction of quantum states~\cite{Bagci_2014,Andrews_2014} to scanning force microscopy~\cite{Halg_2021}, nuclear spin imaging~\cite{Fischer_2019,Kosata_2020}, spin-phonon entanglement~\cite{Karg_2020_light,Thomas_2021_entanglement}, and gravitational wave detection~\cite{Page_2021_gravitational}. These applications profit immensely from cryogenic cooling of the resonator, both due to the lower thermomechanical noise and because the mechanical quality factor $Q$ generally increases with reduced temperatures~\cite{Yuan_2015,fischer_2016}.

    \begin{figure}
    \includegraphics[width=1.02\columnwidth]{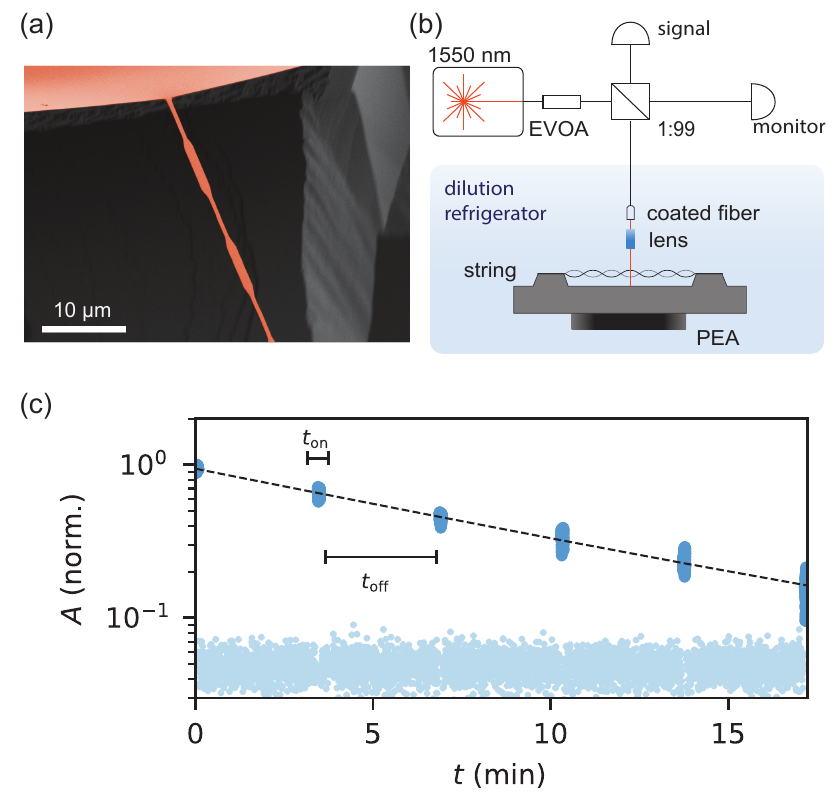}
    \caption{\textbf{Experimental setup and stroboscopic ringdown.}
     (a)~A false-colored SEM image of a string device similar to the ones used in this study. Shown is one clamping point and two unit cells. (b)~Schematic illustration of the setup: A \SI{1550}{\nano\meter} laser is used to measure the displacement of the \ce{Si3N4} beam. An electronic variable optical attenuator (EVOA) is used to switch on and off the optical readout for the stroboscopic measurements. The light reflected from the device interferes with the light reflected at the semi-transparent coated fiber end. The light is focused with a gradient-index (GRIN) lens. The resonator motion is excited via a piezoelectric actuator (PEA) attached to the sample holder. The ac photo detector signal is recorded with a lock-in amplifier. (c)~Stroboscopic ringdown measurement of the localized mode with frequency $f_0=\SI{1.406}{\mega\hertz}$ at $T_\mr{plate}= \SI{46}{\milli\kelvin}$ with $t_{\mr{on}}=\SI{5}{\second}$ and $t_{\mr{off}}=\SI{200}{\second}$. The best fit for this particular ringdown yields $Q = 2.41\pm 0.05\times10^9 $ (95\% confidence interval of the fit). Roughly \SI{120}{\nano\watt} of laser power were focused on the resonator with a duty cycle of $0.02$, yielding an average power of \SI{2.4}{\nano\watt}.
    }
    \label{fig:fig1}
    \end{figure}

Even though \ce{Si3N4} resonators are expected to achieve their best performance at millikelvin temperatures, little is known about their actual properties (intrinsic $Q$, thermal conductivity, Young's modulus, ...) under such conditions~\cite{Yuan_2015,Seis_2021}, in particular in the presence of optical absorption~\cite{fischer_2016,Page_2021_gravitational}. This is in part owed to the technical difficulties added by working in a dry dilution refrigerator, whose vibrations can furthermore add force and frequency noise~\cite{Kalra_2016}. More fundamental, however, is a concern that the interaction with an optical light field will induce heating in the mechanical device, making it difficult to consistently operate at \si{\milli\kelvin} temperatures. This issue could be strong for one-dimensional (string) devices~\cite{Ghadimi_2017,Beccari_2021_hierarchical, Beccari_2022}, whose extreme aspect ratio is expected to lead to inefficient heat conduction.

In this work, we experimentally demonstrate that a  corrugated soft-clamped \ce{Si3N4} string resonator with an aspect ratio of \num{200000} (\SI{4}{\milli\meter} long, \SI{20}{\nano\meter} thick) can be operated at temperatures as low as $46\pm 10$\,mK. We cool the strings in a dry dilution refrigerator with a custom-built vibration isolation and a compact optical interferometer readout that allows us to characterize the resonator decay times during ringdown experiments. Through careful analysis, we infer that the string achieves thermal equilibrium with the sample plate for low optical readout power. Our best device reaches a mean value for the quality factor of $(2.3\pm 0.12) \times 10^9$ at a resonance frequency of $\SI{1.406}{\mega\hertz}$, corresponding to an intrinsic force sensitivity of \SI{9.6}{\zepto\newton\per\sqrt{\hertz}} and a dissipation-limited coherence time of \SI{0.38}{\second}. Our work further shows a strong increase of $Q$ at the lowest measured temperatures, providing fresh insight regarding the fundamental limitations of the coherence of mechanical state.

We study soft-clamped nanostrings made from prestressed \ce{Si3N4}~\cite{Ghadimi_2018}, see Fig.~\ref{fig:fig1}(a). The fabrication procedure is detailed in the Supplemental Material. The devices are mounted in a Leiden CF450 dilution refrigerator with a mixing chamber base temperature of approximately $T_\mr{mc} = \SI{30}{\milli\kelvin}$, corresponding to a sample plate temperature as low as $T_\mr{plate}= \SI{40}{\milli\kelvin}$~\cite{tao_2014_single, heritier_2018_nanoladder, Supplement}. A spring suspension with soft copper braids between the mixing chamber and the sample stage reduces the in-coupling of mechanical noise. Two heaters at the sample stage allow for variations of the sample plate temperature between \num{46} and \SI{194}{\milli\kelvin}.

Mechanical oscillations of the string are measured with a fiber-based optical interferometer, see Fig.~\ref{fig:fig1}(b)~\cite{Rugar_1991,heritier_2018_nanoladder}. A piezoelectric actuator excites mechanical oscillations to the initial amplitude used for ringdown measurements. To reduce optical heating and back-action effects, we apply a stroboscopic measurement scheme~\cite{Ghadimi_2018} as shown in Fig.~\ref{fig:fig1}(c). A variable optical attenuator turns the illumination of the device on and off, enabling ``ringdowns in the dark'' whose progress is only intermittently probed. The photo current from the signal diode is amplified and measured with a lock-in detector. A phase-locked loop (PLL) is used to continuously adjust the local oscillator frequency of the lock-in amplifier to the mechanical resonance frequency during each `on' period.

    \begin{figure}
    \includegraphics[width=1.02\columnwidth]{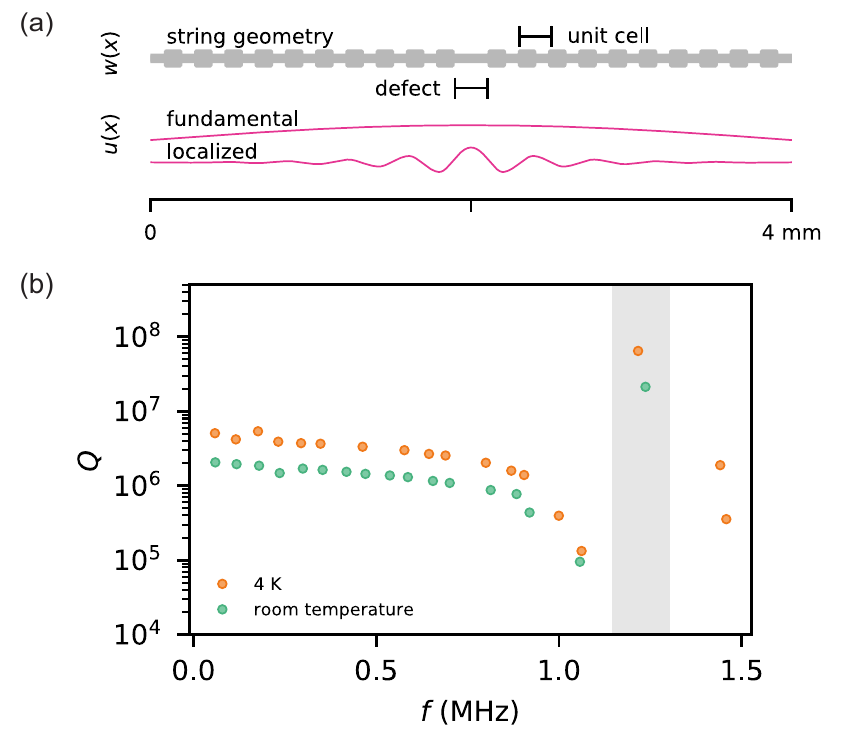}
    \caption{\textbf{String geometry and mode structure of device A.}
    (a)~Shape $w(x)$ of the \SI{4}{\milli\meter} long corrugated beam with 10 unit cells on each side ($n=10$) with a unit cell pitch of \SI{100}{\micro\meter}, filling factor of 50\% ($w_{\mr{min}}=\SI{500}{\nano\meter}$, $w_{\mr{max}}=\SI{1200}{\nano\meter}$), and a defect at the center with a width of \SI{500}{\nano\meter} and length of \SI{120}{\micro\meter}. The corresponding fundamental and localized mode shapes are shown as $u(x)$.
    (b)~Quality factors of various mechanical modes of the corrugated string at room temperature (green) and at \SI{4}{\kelvin} (orange). The string is \SI{100}{\nano\meter} thick and grey shading indicates the extent of the phononic bandgap.}
    \label{fig:fig2}
    \end{figure}

Corrugated nanostrings exhibit a distinctive mode structure which is dictated by the number of unit cells and their geometry. The periodic corrugation creates a phononic crystal (PnC) with a bandgap~\cite{Ghadimi_2017}. This bandgap spatially confines certain flexural modes, whose quality factors are enhanced due to a reduction in the clamping loss, see Fig.~\ref{fig:fig2}(a)~\cite{Ghadimi_2017,Tsaturyan_2017,Fedorov_2019}. For our initial study, we use a \SI{4}{\milli\meter}-long and \SI{100}{\nano\meter}-thick string, with a width varying between $w_{\mr{min}}=\SI{500}{\nano\meter}$ and $w_{\mr{max}}=\SI{1200}{\nano\meter}$ (Device A, see Table.~\ref{tab:tab1}).
For the particular string studied here, the soft-clamped mode has a resonance frequency of $f_{\mr{loc}}=\SI{1.237}{\mega\hertz}$ at room temperature, well inside the phononic bandgap indicated by a shaded region in Fig.~\ref{fig:fig2}(b). The effect of soft-clamping for the in-bandgap mode is clearly visible, since only this mode exceeds $Q>10^7$.

Frequencies as well as quality factors of the mechanical modes are affected by changes in temperature. While the relative changes in frequency are small (e.g. $\delta f_{\mr{loc}}=\SI{20}{\kilo\hertz}$ for a localized mode with $f_{\mr{loc}} = \SI{1.237}{\mega\hertz}$), the quality factors increase significantly. In Fig. \ref{fig:fig2}(b), we observe a consistent increase of the quality factors at \SI{4}{\kelvin} compared to room temperature for all modes.

\begin{table}
\centering
\begin{tabular}{l>{\raggedleft\arraybackslash}p{1.5cm}>{\raggedleft\arraybackslash}p{2cm}>{\raggedleft\arraybackslash}p{2cm}>{\raggedleft\arraybackslash}p{2cm}}
 \rowcolor{beaublue}[1.2\tabcolsep] \multicolumn{2}{l}{} &  Device A &  Device B & Device C\\
 \multicolumn{2}{l}{Length}  & \SI{4}{\milli\meter} & \SI{4}{\milli\meter} &  \SI{4}{\milli\meter}\\
 \multicolumn{2}{l}{Thickness} & \SI{100}{\nano\meter} & \SI{20}{\nano\meter}& \SI{20}{\nano\meter}\\
 \multicolumn{2}{l}{Unit cells} & 10 & 12 & 12\\
 \multicolumn{2}{l}{D-to-UC ratio} & 1.2 & 1.1 & 1.2\\
 \hline
 \multicolumn{2}{l}{$f_\mr{loc}$} & \SI{1.237}{\mega\hertz}& \SI{1.443}{\mega\hertz} & \SI{1.406}{\mega\hertz}\\
 
  \multicolumn{2}{l}{$Q_{300}$ ($\times 10^6$)} & $21$ & $73$& $100$ \\
   \multicolumn{2}{l}{$Q_{4}$ ($\times 10^6$)} & $65$ & $540$ & $730$\\
    \multicolumn{2}{l}{$Q_{\mr{46m}}$ ($\times 10^6$)} &  & $1600$ & $2300$\\

\end{tabular}
\caption{\textbf{List of devices.} The temperature for the $Q_4$ is between \num{4} and \SI{7}{\kelvin}. D-to-UC ratio is the ratio between the length of the defect and the pitch of a unit cell as indicated in Fig.~\ref{fig:fig2}(a).}
\label{tab:tab1}
\end{table}

The stress, and therefore the dissipation dilution factor of pre-stressed resonators, increases for reduced beam cross sections~\cite{Tsaturyan_2017,Ghadimi_2018}. In order to achieve higher quality factors, we repeat the measurements with two $\SI{20}{\nano\meter}$-thick strings. We investigate strings with 12 unit cells, with defect-to-unit-cell (D-to-UC) ratios of 1.1 (device B) and 1.2 (device C). The frequencies of the fundamental modes are $f_1^\mr{B (C)}=\SI{56.6}{\kilo\hertz}\, (\SI{56.3}{\kilo\hertz)}$ and those of the localized mode (mode number 25) are $f_{\mr{loc}}^\mr{B (C)} =\SI{1.443}{\mega\hertz}\,(\SI{1.406}{\mega\hertz)}$ at room temperature.

    \begin{figure}
    \includegraphics[width=1.02\columnwidth]{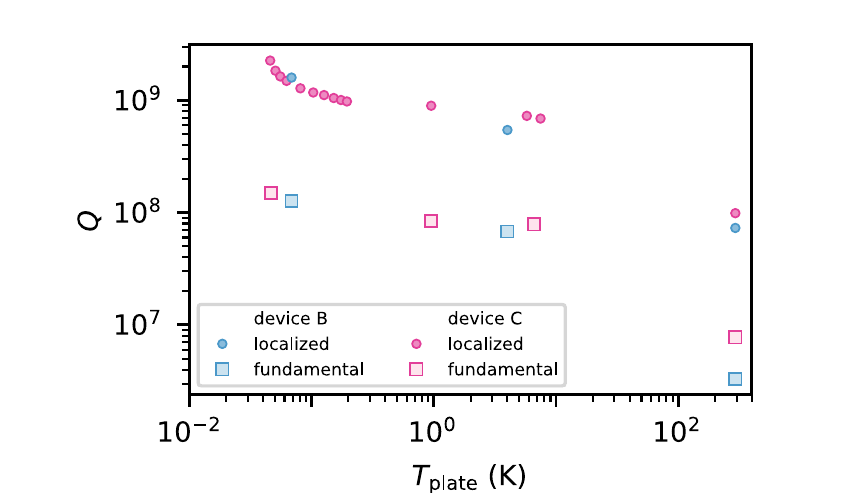}
    \caption{\textbf{Quality factor for various temperatures.}
    Measured $Q$ of two \SI{20}{\nano\meter}-thick strings with $n=12$ unit cells and different defect-to-unit-cell ratio (device B: ratio 1.1, blue, device C: ratio 1.2, red) at various temperatures. For each string, circles and squares correspond to the localized and fundamental modes, respectively.
    }
    \label{fig:fig3}
    \end{figure}

Figure~\ref{fig:fig3} summarizes the $Q$ values we measured for the fundamental and localized modes of the two devices B and C over the full temperature range. The localized modes show a constant and roughly ten-fold higher quality factor compared to the fundamental modes, while the variations between the two devices are small. A reduction of the temperature from room temperature to \SI{4}{\kelvin} leads to an improvement of all quality factor values by a factor $\sim$10. Further cooling to the base temperature of the refrigerator ($T_\mr{mc} = \SI{30}{\milli\kelvin}$) leads to another steep increase of $Q$ for the localized mode of device C reaching up to $Q=2\times 10^9$. The heating experiment has exclusively been conducted with device C. A similar increase has been reported for metal-coated silicon strings~\cite{Maillet_2020} and with silicon nitride membrane resonators~\cite{Yuan_2015,fischer_2016,Page_2021_gravitational,Seis_2021}, but its microscopic origin remains unclear. To confirm the validity of our result, we performed careful studies as a function of average optical power and sample plate temperature that we present in the following. We note that the thermometer installed on the sample plate was not operational during these experiments. The temperature values reported here are the result of a careful calibration study between $T_\mr{plate}$ and $T_\mr{mc}$ that we performed in an additional cooldown. The calibration is shown in the supplemental material~\cite{Supplement}.

    \begin{figure}
    \includegraphics[width=1.02\columnwidth]{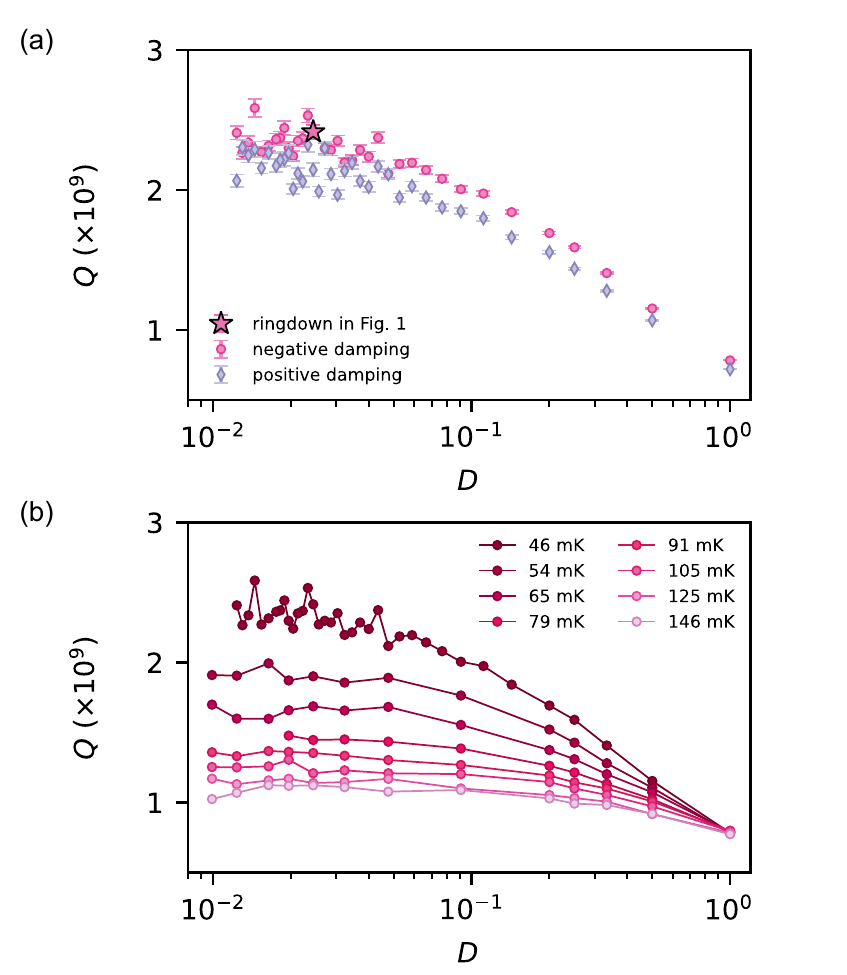}
    \caption{\textbf{Quality factor dependence on laser duty cycle and temperature.}
    (a)~Quality factor of device C as a function of the duty cycle $D$ of the laser in the stroboscopic measurement. We kept $t_\mathrm{on} = \SI{5}{\second}$ and varied $t_\mathrm{off}$. Since the beam waist is larger than the string width, only around \SI{120}{\nano\watt} of the \SI{1.3}{\micro\watt} incident laser power is focused on the string while the laser is on. The interferometer is locked to values that induce either positive and negative damping (cf. SM~\cite{Supplement}). Error bars corresponds to 95\% confidence interval of the best fit.
    (b)~$Q$ of device C for increasing sample plate sensor temperatures $T_\mr{plate}$.}
    \label{fig:fig4}
    \end{figure}

The interaction of the string with the laser beam can induce two effects in the resonator modes: (i)~on the one hand, optical absorption heats up the string above the lattice temperature of the stage it is mounted on. (ii)~On the other hand, radiation pressure forces or photothermal forces act as positive or negative feedback that drive or damp a resonator mode~\cite{Aspelmeyer_2014,Metzger_2004}. To minimize these effects, we employ stroboscopic ringdown measurements, where light illuminates the string only for short periods $t_\mathrm{on}$ between `dark' periods $t_\mathrm{off}$ (see Fig. \ref{fig:fig1}(c)). The duty cycle $D$ of a stroboscopic measurement is defined as $D=t_{\mr{on}}/(t_{\mr{on}}+t_{\mr{off}})$.  We repeat ringdown measurements for positive and negative optical damping to characterize the influence of optomechanical feedback forces (cf. SM~\cite{Supplement} for details regarding the fringe structure).

In Fig.~\ref{fig:fig4}(a), we see that the difference between opposite feedback forces is small. This indicates that the influence of radiation pressure or photo-thermal forces is minor. In contrast, the measured quality factor depends strongly on the duty cycle $D$, i.e., the average heating power arriving on the string surface. The effect is strongest when operating at the base temperature of the cryostat. The quality factor increases as we reduce $D$ and plateaus for $D < 0.03$. This suggests that in this regime (i)~either the string is well thermalized with the sample plate and the laser-induced heating is negligible, or (ii)~$Q$ becomes independent of temperature below $T_\mr{plate}\approx\SI{200}{\milli\kelvin}$.

In Fig.~\ref{fig:fig4}(b), we show repeated measurements of $Q$ as a function of $D$ in the presence of a local heater mounted on the sample plate. Increasing $T_\mr{plate}$ leads to an immediate reduction of $Q$, demonstrating clearly that option (ii) is wrong. We therefore conclude that option (i) is correct: our string resonators do thermalize with the sample plate temperature $T_\mr{plate}$.
%We calibrated in an additional cooldown that the temperature of the sample plate is $T_\mr{plate} = 46\pm 10$\,mK when $T_\mr{mc} = \SI{30}{\milli\kelvin}$~\cite{Supplement}.

    \begin{figure}
    \includegraphics[width=1.02\columnwidth]{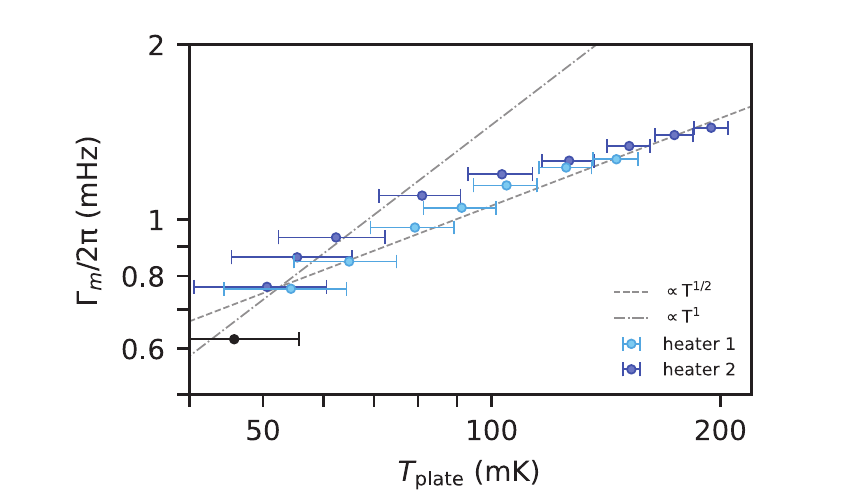}
    \caption{\textbf{Temperature-dependent dissipation coefficient.}
    $\Gamma_m$ measured as a function of $T_\mr{plate}$ with device C. Changes in $T_\mr{plate}$ were induced by two local heaters mounted on the sample stage (heater 1 and 2). The black dot corresponds to the case without heating.
    %Independent calibration data indicate that the sample stage and mixing chamber temperatures rise roughly in proportion (cf. section S3 of \cite{Supplement}).
    The dashed and dashed-dotted grey lines visualize the power laws $\Gamma_m \propto T^\nu$ with $\nu=1/2$~\cite{Seoanez_2008} and $\nu=1$~\cite{Behunin_2016,Hauer_2018} predicted by the STM for one-dimensional resonators. Error bars indicate the calibration uncertainty of \SI{10}{\milli\kelvin} of the sample stage thermometer.
    }% dark blue = closer
    \label{fig:fig5}
    \end{figure}

One possible explanation of the reduced dissipation below \SI{200}{\milli\kelvin} is the non-resonant coupling between ensembles of defects in the material and the motion of the mechanical resonator~\cite{Hauer_2018}. The defects can be modelled as two-level systems (TLS) in an asymmetric double-well potential~\cite{Phillips_1987, Behunin_2016}. The resonator oscillation induces localized strain variations that perturb the potential and couple the defects and phonons~\cite{Grabovskij_2012}.
To compare our data to the standard tunnelling model (STM), we plot $\Gamma_m = 2\mr{\pi} f_{\mr{loc}}/Q$ as a function of $T_\mr{plate}$ in Fig.~\ref{fig:fig5}. Theoretical STM analyses on the role of TLS for nanomechanical resonators predict a power law $\Gamma_m \propto T^{1/2}$~\cite{Seoanez_2008} or $\Gamma_m \propto T$~\cite{Behunin_2016,Hauer_2018} for a quasi-one-dimensional string, with a plateau towards higher temperatures.

The comparison of experimental data to the STM has produced a variety of results in the past. With a range of different materials, measurements of quasi-one-dimensional devices were approximated as $\Gamma_m \propto T^\nu$ with $\nu = 1/3$~\cite{Zolfagharkhani_2005,shim2007micromechanical,Imboden_2009}, $\nu = 2/3$~\cite{Lulla_2013}, or $\nu = 1$~\cite{Hauer_2018,Maillet_2020}. A saturation of $\Gamma_m$ towards low temperatures was accounted for by the form $\Gamma_m \propto (1+(T/T_0)^\nu)$, with the free parameters $T_0 = \SI{0.3}{\kelvin}$ and $\nu = 1.6$~\cite{tao_2014_single,Seoanez_2008}. Our data shows a plateau of the dissipation above $T_\mr{plate}\approx \SI{150}{\milli\kelvin}$ but no saturation at the lower end of the accessible temperature range.

The implications of our measurements remain to be investigated. In future work, we will study the possible presence of surface adsorbates that can introduce additional damping~\cite{Heritier_2021}, and their partial disorption with strong local heaters~\cite{Martinez_2013}.

Looking forward, high $Q$ and low device temperatures are essential when using mechanical resonators for the detection of small forces. The nanostrings investigated in this work have a thermal noise-limited, single-sided force sensitivity estimated as~\cite{Saulson_1990}
\begin{equation}
    \sqrt{S_f} =  \sqrt{4k_\mr{B}T_\mr{plate} m_{\mr{eff}}\Gamma_m}=\SI{9.6}{\zepto\newton\per\sqrt\hertz}
\end{equation}
where $k_\text{B}$ is the Boltzmann constant, $T_\mr{plate} = \SI{46}{\milli\kelvin}$ is the device temperature, $\Gamma_m = 2\mr{\pi} f_0/Q$ is the dissipation coefficient for a resonance frequency $f_0$, and the effective mass $m_{\text{eff}}=\SI{9.3}{\pico\gram}$ is extracted from numerically solving the one-dimensional Euler-Bernoulli equation~\cite{Fedorov_2019}. Our best device therefore attains a force sensitivity similar to that of a single carbon nanotube~\cite{Moser_2013,deBonis_2018_ultrasensitive}, in spite of the roughly $10^6$ times larger mass. In contrast to carbon nanotubes with a diameter of $\approx\SI{1}{\nano\meter}$, our top-down patterned \ce{Si3N4} strings are large enough to envisage mounting a molecular sample on them. This provides an interesting perspective for ultrasensitive force detection experiments, for instance in the context of nanoscale magnetic resonance imaging~\cite{Poggio_2010,Nichol_2013,Grob_2019,Kosata_2020}. A side-by-side comparison with the performance a state-of-the-art cantilever sensor confirms this assessment (cf. SM~\cite{Supplement}).

Low temperatures and high quality factors are also important for increasing the quantum coherence time of the resonator's oscillation states~\cite{Rossi_2018,Seis_2021}. Dissipation imposes a limit to the coherence time, yielding in our case
\begin{equation}
    \gamma_\mathrm{diss}^{-1} \approx \frac{1}{n\Gamma_m} =  \frac{Q \hbar}{k_\mr{B} T_\mr{plate}} = \SI{0.38}{\second}
\end{equation}
where $n\approx 1000$ is the phonon occupation number of the resonator mode and $\hbar$ is the reduced Planck constant. We emphasize, however, that frequency fluctuations~\cite{Fong_2012} can significantly reduce this coherence time in practice. The characterization of frequency fluctuations in these corrugated string devices is an important topic to be explored in future studies.

A second crucial objective is to reduce the detection noise when reading out the nanomechanical oscillation. The simple interferometer used in the current setup is not optimized for string devices and produces significant additional readout noise, making the detection of thermomechanical force noise impossible. Integrated optomechanical readout~\cite{Schilling_2016,Guo_2019} can mitigate this problem. Increasing the vacuum optomechanical coupling rate $g_0$~\cite{Aspelmeyer_2014} will allow better detection sensitivity for the same added heating. This is possible because the optomechanical cooperativity scales as $G^2$, while absorption of light, in a rough approximation, is estimated to increases only linearly with $G$~\cite{wilson2015measurement,ghadimi2018ultra}.

In summary, we cooled and operated a \ce{Si3N4} string to a bath temperature of \SI{46}{\milli\kelvin}, where it achieved a force sensitivity of $\SI{9.6}{\zepto\newton\per\sqrt{\hertz}}$ and a dissipation-limited coherence time of $\SI{0.38}{\second}$. We measure mechanical $Q$ as high as $2.3\times10^9$, more than $20\times$ improvement compared to room temperature. We observe a sudden increase of $Q$ at the lowest temperatures of our dilution refrigerator, implying that further improvements should be achieved with a larger cooling power at base temperature and with better thermalization of the sample plate. The microscopic origin of the damping and decoherence in our devices is not fully understood. However, the isolated nature of these localized modes provides an unique avenue for further work. Due to their simple geometry and strong isolation from environmental influences, nanomechanical strings offer an ideal system to test model predictions.

\section*{Acknowledgments}

This work was supported by the Swiss National Science Foundation (CRSII5\_177198/1) and an ETH Zurich Research Grant (ETH-51 19-2).

\providecommand{\noopsort}[1]{}\providecommand{\singleletter}[1]{#1}%

\end{document}

\title{\textbf{Supplemental Material for:}\\ Soft-clamped silicon nitride string resonators at millikelvin temperatures}

\author{Thomas Gisler}
\affiliation{\affilETH}
\author{Mohamed Helal}
\affiliation{\affilETH}
\author{Deividas Sabonis}
\affiliation{\affilETH}
\author{Urs Grob}
\affiliation{\affilETH}
\author{Martin H\'eritier}
\affiliation{\affilETH}
\author{Christian L. Degen}
\affiliation{\affilETH}
\author{Amir H. Ghadimi}
\email[Corresponding author: ]{amir.ghadimi@csem.ch}
\affiliation{\affilCSEM}
\author{Alexander Eichler}
\email[Corresponding author: ]{eichlera@ethz.ch}
\affiliation{\affilETH}

\maketitle

\spacing{1.5}
\section{Optical fringes}
\begin{figure}
    \includegraphics[width=\textwidth]{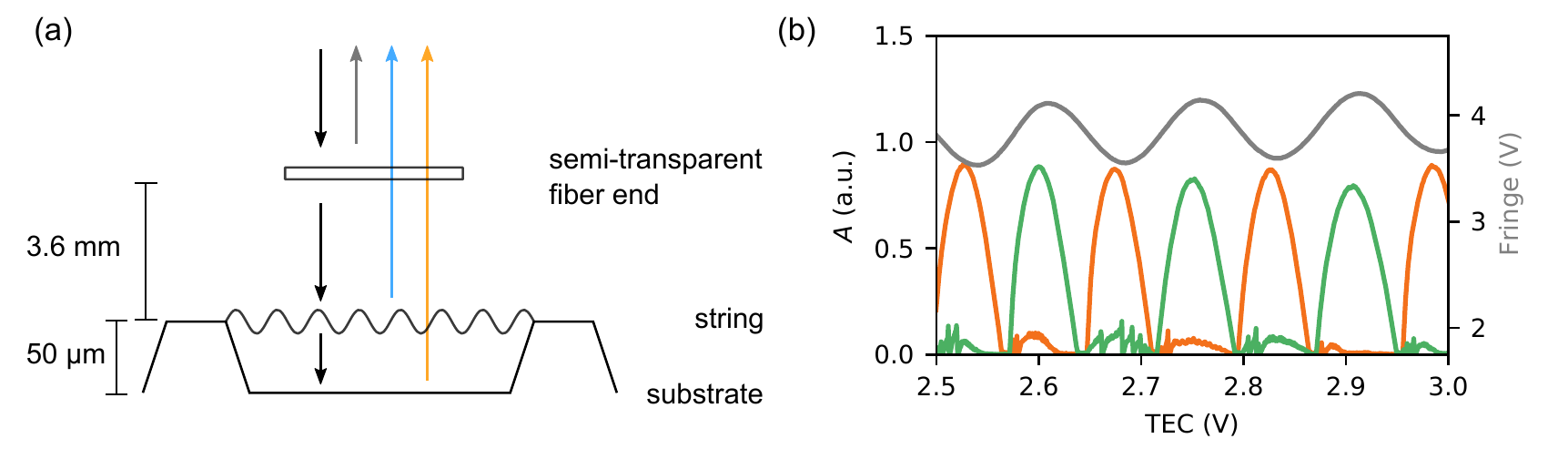} 
    \caption{\textbf{Interferometric detection method.} (a)~Schematic of the optical interferometer, with light reflected from the fiber end (grey), the string (blue) and the substrate (orange). (b)~Sweeping the TEC voltage (corresponds to laser wavelength) and measuring the optical fringe (grey - right y-axis) and the amplitude of the fundamental mode with two different PLL setpoint phases $\phi_\mr{PLL}$ (orange: $\phi_\mr{PLL}=\phi_0$, green: $\phi_\mr{PLL}=\phi_0+\mr{\pi}$, left y-axis). 
    }
    \label{fig:SIFigFringe}
    \end{figure}

Tuning the wavelength by changing the temperature of the laser diode allows to move within the optical fringe. For this we apply a voltage to the thermoelectric cooler (TEC) pins of the butterfly laser diode. While the motion of the string is encoded in the interference of the light reflected from the string resonator and the coated fiber end, the main fraction of the reflected light comes from the sample chip substrate, see Fig.~\ref{fig:SIFigFringe}(a). This results in the optical fringe shown in Fig.~\ref{fig:SIFigFringe}(b). Using a phase-lock loop (PLL) to drive and measure the displacement of the fundamental mode of the string - while sweeping the TEC voltage - reveals the position within the fringe where the largest signal is obtained. The TEC voltage sweep is performed for two different PLL setpoint phases shifted by $\mr{\pi}$ (orange and green in Fig.~\ref{fig:SIFigFringe}(b)). The mechanical signal is not maximized at the steepest slope of the optical fringe, as is typically observed with this interferometer setup~\cite{heritier_2018_nanoladder}. Instead, the maximum position cannot be predicted from the fringe signal and must be measured directly as a function of the TEC voltage, as described above. The laser wavelength is then locked to this point using a PID feedback. This point corresponds to where the phase light reflected from the fiber end and the string are shifted by $\pm\mr{\pi}$. The maxima of the orange and green signals correspond to maximum positive and negative optomechanical damping, respectively.

\section{Repeatability of ringdown measurements}
\begin{figure}
    \includegraphics[width=\textwidth]{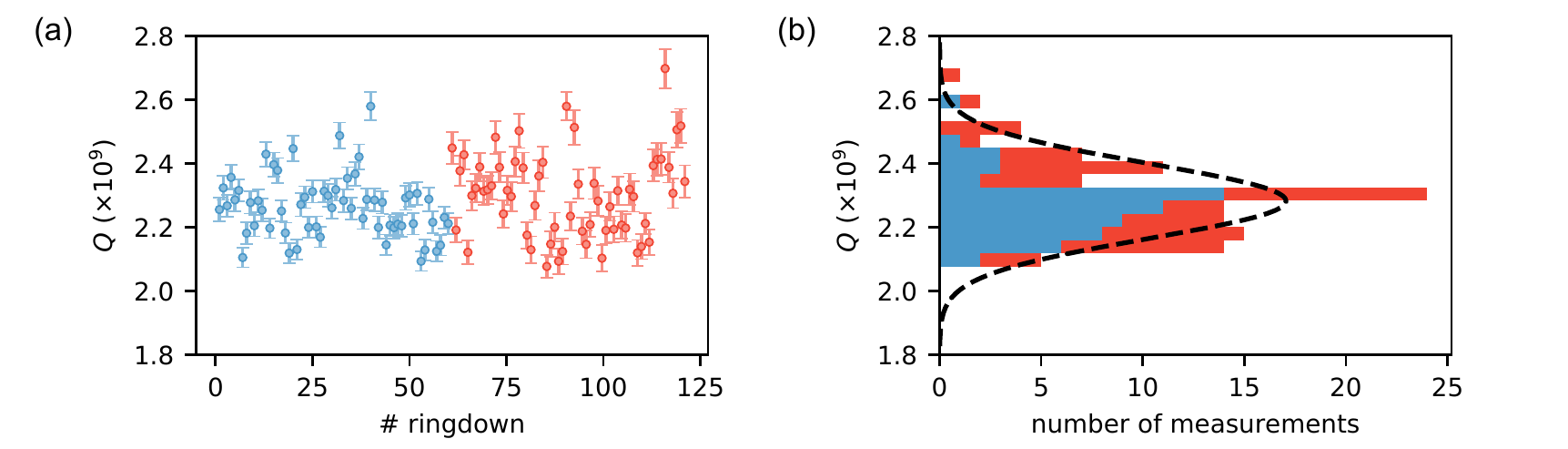} 
    \caption{\textbf{Repeatability of measured $Q$.}
    Repetition of ringdown measurements with $t_\mr{on} = \SI{5}{\second}$ for two different ``laser off'' periods ($t^\mr{blue}_\mr{off}=\SI{150}{\second}$ and $t^\mr{red}_\mr{off}=\SI{250}{\second}$). (a)~Measured $Q$ and (b)~histogram of these values together with a Gaussian fit with mean value \num{2.282e9} and standard deviation \num{0.116e9}.
    }
    \label{fig:SIFigRep}
    \end{figure}

\begin{figure}
    \includegraphics[width=\textwidth]{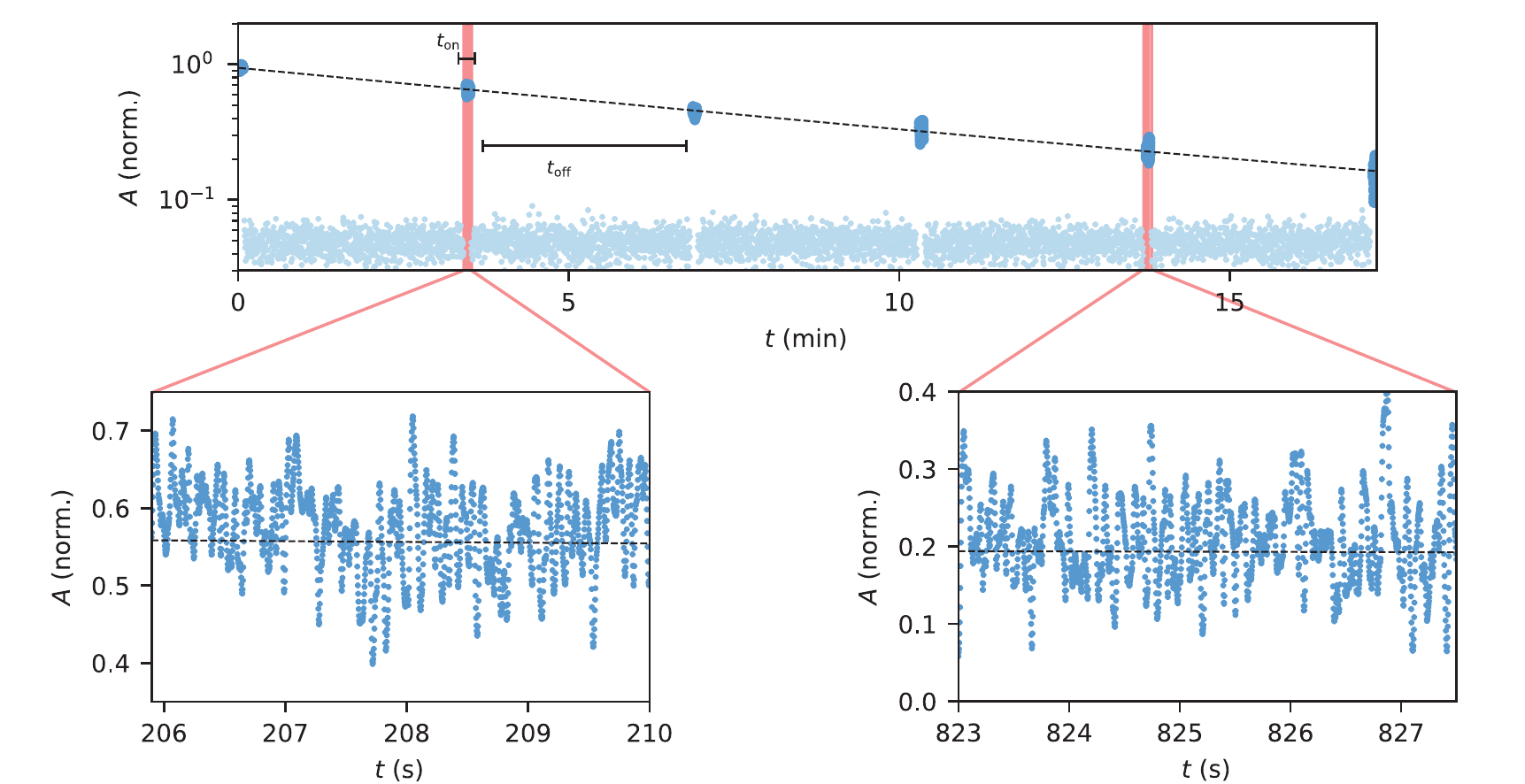} 
    \caption{\textbf{Zoom ringdown}
    The same ringdown as in Fig.~1(c) presented with zoomed details. The large spread of data points is due to detector noise.
    }
    \label{fig:SIFigRingdown}
    \end{figure}

In order to verify the measured $Q$ we repeatedly conducted ringdown measurements with the same settings. In Fig.~\ref{fig:SIFigRep}(a-b) we collect the extracted $Q$ values of the localized mode for $t_\mr{off}=\SI{150}{\second}$ and \SI{250}{\second} of device B at $T_{plate}=\SI{48}{\milli\kelvin}$ and $t_\mr{on}=\SI{5}{\second}$. The mean of the measured values is \num{2.282e9} with a standard deviation of \num{0.116e9}. This mean value for $Q$ is used to calculate the intrinsic force sensitivity and thermal decoherence time in the main text. The zoomed data in Fig.~\ref{fig:SIFigRingdown} demonstrate the large spread of data points due to the low-pass filtered detector noise.

\section{Variable temperature measurements}
\begin{figure*}
    \includegraphics[width=\textwidth]{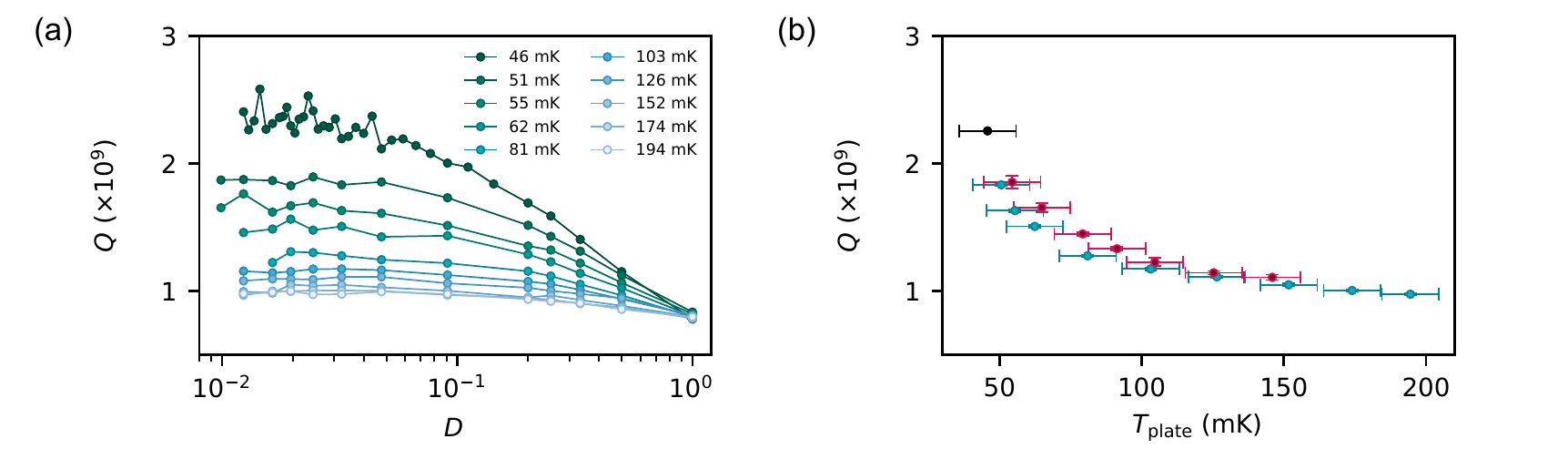} % size for 2column
    \caption{\textbf{Comparing the different heaters.}
    (a)~Driven PEA with frequency {\SI{1.5}{\mega\hertz}} with driving amplitudes between \SI{50}{\milli\volt} and \SI{200}{\milli\volt}. The temperatures in the legend correspond to $T_\mr{plate}$. Note that the top line, corresponding to the unheated string, is identical to the measurement in the main text. {(b)}~Comparison of $Q$ as a function of $T_\mr{plate}$ for the two heaters (blue closer to the sample than purple).
    }
    \label{fig:SIFigDC}
    \end{figure*}

We used two different heaters to locally change the temperature of the nanomechanical resonator. For each temperature $T_\mr{plate}$ (extracted from the mixing chamber temperature, see Fig.~\ref{fig:SIFigTemperature}) we extracted $Q$ for different duty cycles $D$. In the main text, data are shown for a resistive heater mounted a few centimeters from the sample on the sample plate. In Fig.~\ref{fig:SIFigDC}(a) we supply the results for the second heater, consisting of the piezoelectric actuator (PEA) mounted closer to the string chip. Here, a temperature variation was induced by applying a high-frequency signal $f_\mr{heat}=\SI{1.5}{\mega\hertz}$ to the PEA, carefully avoiding any mechanical resonance frequency. In Fig.~\ref{fig:SIFigDC}(b) we compare the results of the two heating methods and obtain very similar $Q$ values.

\section{Calibration of sample plate temperature}
\begin{figure}
    \includegraphics[width=\textwidth]{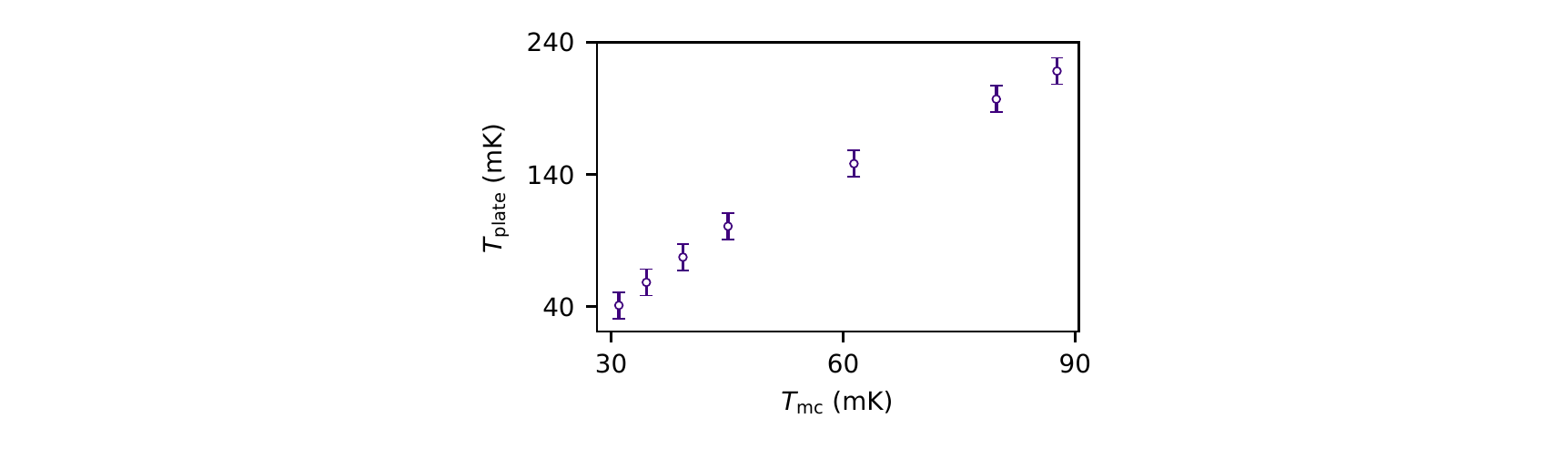} % size for 2column
    \caption{\textbf{Calibration of sample plate temperature.}
    Sample plate temperature $T_\mr{plate}$ as function of the mixing chamber temperature $T_\mr{mc}$ for variable power delivered to a heater on the sample plate. The error bars correspond to the calibration uncertainty (\SI{10}{\milli\kelvin}) of the RuO$_2$ sensor.}
    \label{fig:SIFigTemperature}
    \end{figure}

During the measurements presented in the main text, the thermometer mounted on the sample plate could not be used due to a wiring problem. We later repeated the cooldown (without a mechanical resonator device) to calibrate the temperature difference between the mixing chamber plate and the sample plate. All temperatures in the main text are stated after the conversion using the calibration measurement shown in Fig.~\ref{fig:SIFigTemperature}. We infer a base temperature of $\SI{41}{\milli\kelvin}$ on the sample plate with a systematic uncertainty of \SI{10}{\milli\kelvin} stemming from the sensor calibration. The measurement was performed with a lock-in amplifier at a bias current of roughly \SI{1}{\nano\ampere} applied with a local oscillator frequency of \SI{33}{\hertz}. The power dissipated over the temperature sensor was below \SI{1}{\pico\watt}.

\section{Silicon nitride strings}
\subsection{Fabrication}

The devices used in this work were designed and fabricated at CSEM based on the previous work by authors \cite{Ghadimi_2018,ghadimi2018ultra}. Briefly, the fabrication process flow starts with a low-pressure chemical vapor deposition (LPCVD) of high-stress \ce{Si3N4} on a Si substrate. This is followed by electron beam (ebeam) lithography  of the corrugated string geometry on top of \ce{Si3N4} layer using a flowable oxide (FOX-16 \textregistered) resist (2). Structures are then transferred to the \ce{Si3N4} layer by reactive ion etching (RIE). Finally, the mechanical beams are released from the underlying Si substrate in potassium hydroxide (KOH) bath and drying using a critical point dryer (CPD). The complete process flow is considerably more complex and includes several intermediate steps to ensure the prevention of collapsing of released nanobeams due to their extreme aspect ratios. The detailed process flow can be found in \cite{Ghadimi_2018,ghadimi2018ultra}.

\subsection{Geometrical shape of strings}
\begin{figure}
    \includegraphics[width=\columnwidth]{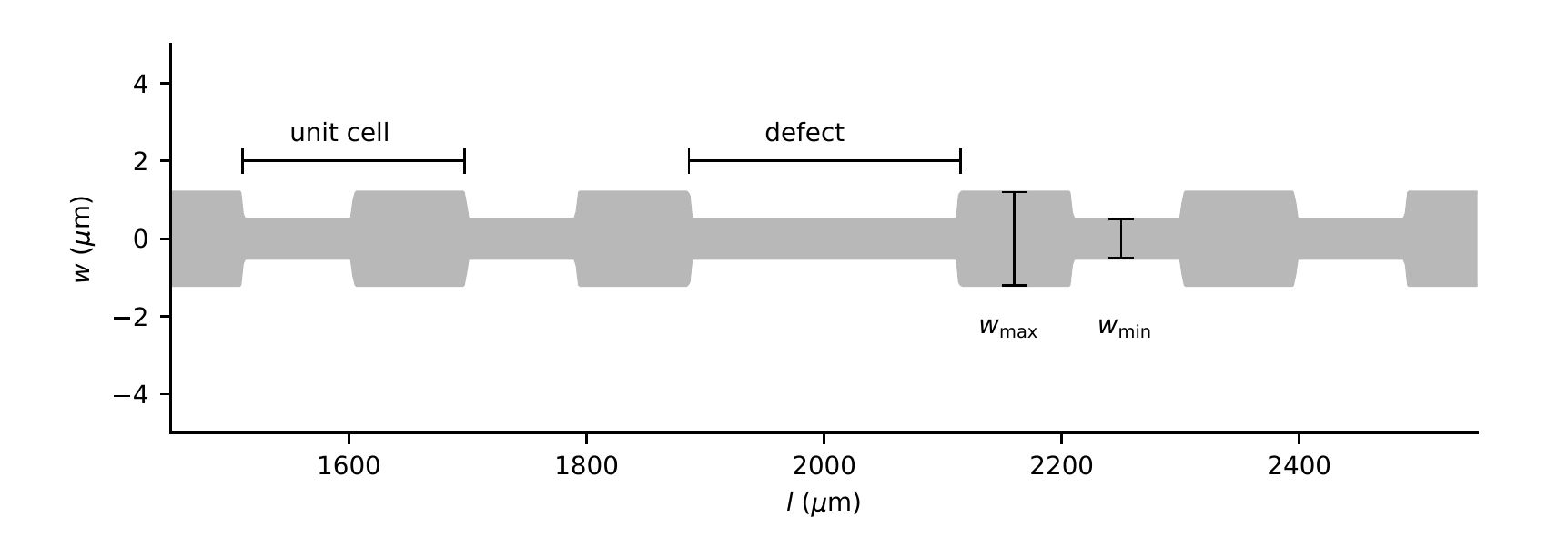} 
    \caption{\textbf{Nanostring geometry.}
    Geometrical shape of the silicon nitride string with $w_\mr{min}=\SI{500}{\nano\meter}$ and $w_\mr{max}=\SI{1200}{\nano\meter}$ and a length of \SI{4}{\milli\meter}.
    }
    \label{fig:SIFigGeo}
    \end{figure}

In this work, we used different corrugated \ce{Si3N4} strings with similar geometries. Figure \ref{fig:SIFigGeo} shows the area around the defect in the center of such a string. Device B and C have the same thickness, size and number of unit cell but their defect-to-unit-cell ratio differs (Device B 1.1 and device C 1.2). 

\subsection{Force noise evaluation for nanoscale MRI}
\begin{figure}
    \includegraphics[width=\columnwidth]{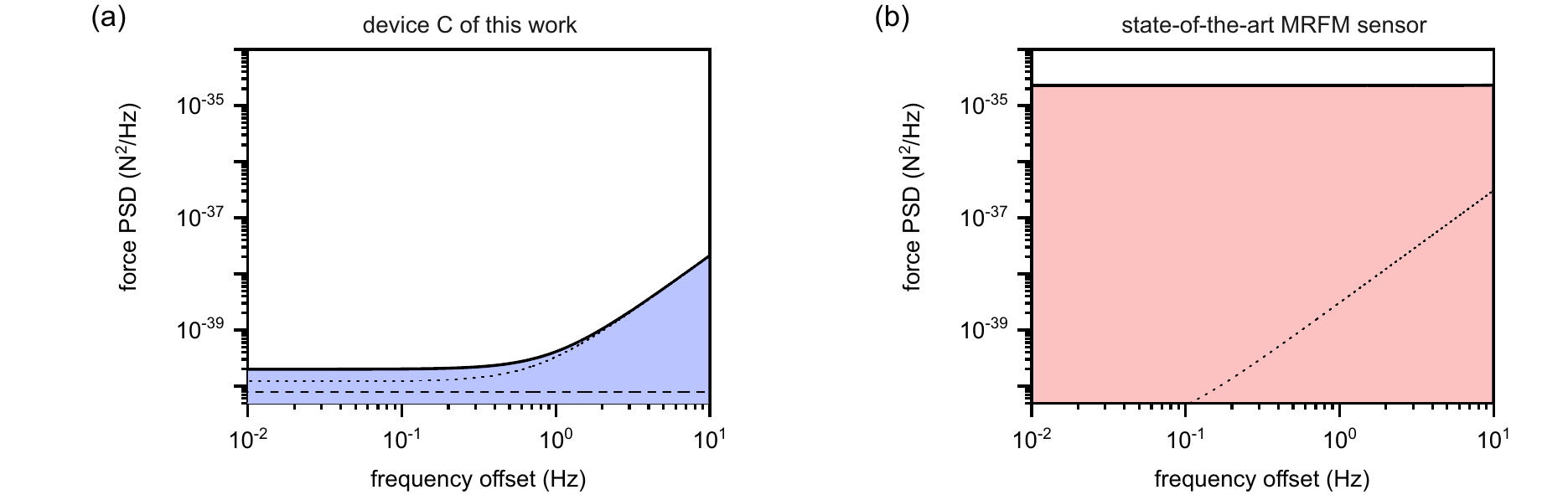} 
    \caption{\textbf{Force sensitivity comparison as a function of the detuning from the resonance frequency $f_0$.}
    (a)~Estimated contributions to the force noise power spectral density of device C measured in this work at a conservative temperature of \SI{80}{\milli\kelvin}. We assume that an on-chip cavity readout can reach a displacement sensitivity of \SI{2e-28}{\meter\squared\per\hertz} without significant additional heating~\cite{Schilling_2016,Guo_2019}. For the magnetic field gradient produced by a coated AFM tip at a distance of $10$ to  \SI{30}{\nano\meter}, we use a conservative value of \SI{1e7}{\tesla\per\meter}. (b)~Cantilever device used in state-of-the-art NanoMRI measurements with a mass of \SI{1e-13}{\pico\gram}, a resonance frequency of \SI{3.5}{\kilo\hertz} and a quality factor of $25000$, (manuscript in preparation, similar performance as in Ref.~\cite{Grob_2019}). This sensor is operated at \SI{4.7}{\kelvin} and no on-chip cavity readout is possible in the present setup, limiting the displacement sensitivity to \SI{4e-24}{\meter\squared\per\hertz}. The magnetic field gradient produced by the nanomagnet in this experiment was found to have a maximum value of \SI{2.8e6}{\tesla\per\meter}. In both panels, a dashed black line is the thermal noise (including the zero-point fluctuations), a dotted black line is the sum of quantum back-action and detector noise, and a thick black line and the colored area indicate the total force noise. The noise of the cantilever device in (b) is dominated by the thermomechanical noise contribution, such that the solid and dashed black lines overlap.
    }
    \label{fig:force_PSD_calculations}
    \end{figure}

The devices used in this work are promising candidates for force sensing in the zeptonewton range. Such a force sensitivity is crucial, for instance, for nanoscale magnetic resonance microscopy (NanoMRI) and for coherent spin-mechanics coupling. In Fig.~\ref{fig:force_PSD_calculations}, we compare the predicted performance of device C measured in this work with a state-of-the-art cantilever sensor used for our most recent NanoMRI demonstration~\cite{krass2022force}, which is comparable to that used in previous experiments~\cite{Grob_2019}. The only modification that we additionally assume for the string device in Fig.~\ref{fig:force_PSD_calculations}(a) is the implementation of an on-chip cavity readout with a displacement sensitivity of \SI{2e-28}{\meter\squared\per\hertz}, as discussed in the main text~\cite{Schilling_2016,Guo_2019}. For a bandwidth of \SI{1}{\hertz}, the present string device has a total force noise standard deviation of \SI{1.7e-20}{\newton}, corresponding to the force generated by $0.0.1$ hydrogen atoms in the magnetic field gradient predicted from finite element calculations. The cantilever device in Fig.~\ref{fig:force_PSD_calculations}(b) achieves a force noise standard deviation of \SI{3.4e-18}{\newton} in the same bandwidth, resulting in a theoretical sensitivity of roughly $7400$ hydrogen atoms. The string resonator is therefore expected to produce highly superior results, allowing the detection of the forces generated by individual nuclei.

%\section{Two-level system}
%Defects at low temperature modeled as ensemble of two-level systems.

%    \begin{figure}
%    \includegraphics[width=0.52\columnwidth]{SIFigTLS.pdf} 
%    \caption{\textbf{Dissipation at low temperature.}
%    Fitting the slope of the dissipation for the different heating methods. The piezo heater is closer to the device, which means we underestimate the temperature. For the resistive heater we the opposite is the case.
%    }
%    \label{fig:SIFig3}
%    \end{figure}

%